\documentclass[%
 reprint,
superscriptaddress,
 amsmath,amssymb,
aps,
]{revtex4-1}
\usepackage{graphicx}
\usepackage{amssymb}
\usepackage{dcolumn}
\usepackage{bm}
\usepackage{xcolor}
\usepackage{todonotes}
\usepackage{hyperref}
\hypersetup{
    colorlinks=true,
    linkcolor=blue,
    filecolor=magenta,      
    urlcolor=red,
    citecolor=blue,
}
\newcommand{\blue}{\textcolor{blue}}

\usepackage{xcolor}

\usepackage{soul}

\begin{document}

\title{Charge redistribution, charge order and plasmon in La$_{2-x}$Sr$_{x}$CuO$_{4}$/La$_{2}$CuO$_{4}$ superlattices}

\author{Qizhi Li}\thanks{These authors contributed equally to this work.}
\affiliation{International Center for Quantum Materials, School of Physics, Peking University, Beijing 100871, China}

\author{Lele Ju}\thanks{These authors contributed equally to this work.}
\affiliation{Interdisciplinary Center for Quantum Information, State Key Laboratory of Modern Optical Instrumentation, and Zhejiang Province Key Laboratory of Quantum Technology and Device, Department of Physics, Zhejiang University, Hangzhou 310027, China}

\author{Hsiaoyu Huang}
\affiliation{National Synchrotron Radiation Research Center, Hsinchu 30076, Taiwan}

\author{Yuxuan Zhang}
\affiliation{National Center for Electron Microscopy in Beijing, Key Laboratory of Advanced Materials (MOE), State Key Laboratory of New Ceramics and Fine Processing, School of Materials Science and Engineering, Tsinghua University, Beijing 100084, China}

\author{Changwei Zou}
\affiliation{International Center for Quantum Materials, School of Physics, Peking University, Beijing 100871, China}

\author{Tianshuang Ren}
\affiliation{Interdisciplinary Center for Quantum Information, State Key Laboratory of Modern Optical Instrumentation, and Zhejiang Province Key Laboratory of Quantum Technology and Device, Department of Physics, Zhejiang University, Hangzhou 310027, China}

\author{A. Singh}
\affiliation{National Synchrotron Radiation Research Center, Hsinchu 30076, Taiwan}
\affiliation{Department of Physics and Astrophysics, University of Delhi, New Delhi 110007, India}

\author{Shilong Zhang}
\affiliation{International Center for Quantum Materials, School of Physics, Peking University, Beijing 100871, China}

\author{Qingzheng Qiu}
\affiliation{International Center for Quantum Materials, School of Physics, Peking University, Beijing 100871, China}

\author{Qian Xiao}
\affiliation{International Center for Quantum Materials, School of Physics, Peking University, Beijing 100871, China}

\author{Di-Jing Huang}
\affiliation{National Synchrotron Radiation Research Center, Hsinchu 30076, Taiwan}

\author{Yanwu Xie}
\affiliation{Interdisciplinary Center for Quantum Information, State Key Laboratory of Modern Optical Instrumentation, and Zhejiang Province Key Laboratory of Quantum Technology and Device, Department of Physics, Zhejiang University, Hangzhou 310027, China}

\author{Zhen Chen}
\affiliation{Beijing National Laboratory for Condensed Matter Physics, Institute of Physics, Chinese Academy of Sciences, Beijing 100190, China}

\author{Yingying Peng}
\email{yingying.peng@pku.edu.cn}
\affiliation{International Center for Quantum Materials, School of Physics, Peking University, Beijing 100871, China}
\affiliation{Collaborative Innovation Center of Quantum Matter, Beijing 100871, China}

\date{\today}

\begin{abstract}

Interfacial superconductors have the potential to revolutionize electronics, quantum computing, and fundamental physics due to their enhanced superconducting properties and ability to create new types of superconductors. The emergence of superconductivity at the interface of La$_{2-x}$Sr$_{x}$CuO$_{4}$/La$_{2}$CuO$_{4}$ (LSCO/LCO), with a T$_c$ enhancement of $\sim$ 10 K compared to the La$_{2-x}$Sr$_{x}$CuO$_{4}$ bulk
single crystals, provides an exciting opportunity to study quantum phenomena in reduced dimensions. To investigate the carrier distribution and excitations in interfacial superconductors, we combine O K-edge resonant inelastic X-ray scattering and atomic-resolved scanning transmission electron microscopy measurements to study La$_{2-x}$Sr$_{x}$CuO$_{4}$/La$_{2}$CuO$_{4}$ superlattices (x=0.15, 0.45) and bulk La$_{1.55}$Sr$_{0.45}$CuO$_{4}$ films. We find direct evidence of charge redistribution, charge order and plasmon in LSCO/LCO superlattices. Notably, the observed behaviors of charge order and plasmon deviate from the anticipated properties of individual constituents or the average doping level of the superlattice. Instead, they conform harmoniously to the effective doping, a critical parameter governed by the T$_c$ of interfacial superconductors.

\end{abstract}

\maketitle
\section{Introduction}

The role of quasi-two dimensionality is believed to be crucial in unconventional superconductors, such as the CuO$_2$ planes in cuprates and the FeSe planes in iron-based superconductors. To achieve the two-dimensional limit, various methods have been employed, including intercalation to separate superconducting layers \cite{inter2010,FeSe11111}, exfoliation of bulk materials to produce single or few-layer sheets \cite{pnas1uc,monoBi2212}, and thin-films deposition onto substrates to create interfacial superconductors \cite{KTOSTOSC,interfacialSC,Juleleprb}. Interface superconductivity has been achieved in a range of materials, like LaAlO$_3$/SrTiO$_3$ \cite{LAOSTOSC}, KTaO$_3$/SrTiO$_3$ \cite{KTOSTOSC} and FeSe/SrTiO$_3$ \cite{Wang_2012}. The construction of interfacial superconductivity offers several advantages: Firstly, the superconducting plane is well-protected at the interface, which can increase the stability and durability of materials. Secondly, unique electronic properties of the interface can lead to higher critical temperatures (T$_c$) than in single-phase compositions. For instance, the T$_{c,max}$ of FeSe is enhanced from 8 K in bulk to nearly 70 K in FeSe/SrTiO$_3$ interface \cite{FeSenature,Song2019}. Finally, interfacial superconductors hold the potential for developing new electronic devices and technologies \cite{ISCdevice}.

The La$_{1.55}$Sr$_{0.45}$CuO$_{4}$/La$_{2}$CuO$_{4}$ bilayer (LSCO/LCO) is a promising platform for studying interfacial superconductivity (SC).
The T$_c$ of LSCO/LCO can exceed 50 K, which is higher than the 40 K of optimal doped bulk LSCO single crystals \cite{interfacialSC}. 
The bilayers comprise two non-superconducting compositions, which enables the separate investigation of interfacial SC from bulk SC. Logvenov $et~al.$ successfully doped Zn atoms into a target CuO$_2$ layer using the molecular beam epitaxy (MBE) method and suppressed the local SC. They determined that the interface superconductivity arises from a single CuO$_2$ plane located in the second layer adjacent to the interface \cite{dopingZn}. Resonant soft X-ray scattering (RSXS) study on La$_{1.64}$Sr$_{0.36}$CuO$_{4}$/La$_{2}$CuO$_{4}$ (T$_c$ = 38 K) superlattices also revealed a redistribution of conducting holes from LSCO to the LCO layers, inducing LCO layers to be optimally doped for superconductivity \cite{LSCO/LCOREXS}. 
Notably, the T$_c$ of the LSCO/LCO interface has shown sequence-dependent behavior \cite{interfacialSC}, and the phase diagram of bilayers has displayed two separate superconducting domes \cite{XieyanwuPLD}. These results indicate that the observed interfacial SC in LSCO/LCO bilayers cannot be solely attributed to carrier reconstruction and requires further study. 

Competing orders and elementary excitations such as charge density wave (CDW) and plasmon are closely correlated with SC in cuprates. CDW is ubiquitous in cuprates family and has been a topic of extensive research in recent years \cite{tranquada1995evidence,abbamonte2005spatially,Ghiringhelli2012,chang2012direct,YYPBi2201,GiacomoScienceCDF}. For example, RSXS experiments on underdoped YBa$_2$Cu$_3$O$_{6.67}$ have demonstrated that CDW is suppressed in the superconducting state and becomes stronger when a magnetic field is applied to suppress SC, indicating a competitive relationship between CDW and SC \cite{chang2012direct}. 
Meanwhile, plasmon, which arises from Coulomb interactions between the two-dimensional electron gas of neighboring CuO$_2$ planes \cite{FETTER1,FETTER2}, has been observed in both electron-doped \cite{LCCOplasma,Gappedplasmon} and hole-doped cuprates \cite{plasmonZhouKJ,Hepting_2023}. Recently, the plasmon has drawn attention due to its potential role in mediating the formation of Cooper pairs \cite{LCCOplasma} or contributing to approximately 20\% of T$_c$ \cite{plasmonmediateCP,plasmon20}. 
The LSCO/LCO superlattices are composed of CuO$_2$ planes with separate dopings. It remains unknown how CDW and plasmon evolve in the LSCO/LCO interfacial superconductors, whether they adhere to the behaviors of separate constituents or exhibit variations owing to the intricate interplay among the CuO$_2$ planes.

In this study, we combine scanning transmission electron microscopy (STEM) and resonant inelastic X-ray scattering (RIXS) to study the carrier distribution and elementary excitations in La$_{2-x}$Sr$_{x}$CuO$_{4}$/La$_{2}$CuO$_{4}$ interfacial superconductors. By utilizing electron energy loss spectroscopy (EELS) measurements, we find that the hole density displays evident charge redistribution along the depth direction, which differs from the La/Sr concentration distribution in the La$_{1.55}$Sr$_{0.45}$CuO$_{4}$/La$_{2}$CuO$_{4}$ superlattice film (T$_c$ $\simeq$ 30 K). Oxygen $K$-edge resonant inelastic X-ray scattering (RIXS) was employed to investigate La$_{1.55}$Sr$_{0.45}$CuO$_{4}$/La$_{2}$CuO$_{4}$ and La$_{1.85}$Sr$_{0.15}$CuO$_{4}$/La$_{2}$CuO$_{4}$ (T$_c$ $\simeq$ 20 K) superlattices, as well as a La$_{1.55}$Sr$_{0.45}$CuO$_{4}$ metallic film for comparison.  
We reveal charge order, phonons and plasmon in LSCO/LCO superlattices, which deviate from the properties of the individual constituents, but conform to the effective doping determined by interfacial superconductivity. Our findings uncover the unique properties of LSCO/LCO superlattices and provide new insights on interfacial superconductors.

\section{Materials and methods}

\subsection{Sample preparation and characterization}

The La$_{2-x}$Sr$_x$CuO$_4$ films were grown on LaSrAlO$_4$ (001) single-crystal substrates using pulsed laser deposition (PLD). Before deposition,  the targets were sintered in Muffle furnaces with nominally stoichiometric compositions. The substrates were pre-treated in situ at 800 $^0$C for 20 minutes, under the oxygen pressure of 1$\times$10$^{-4}$ mbar. All the films were deposited at 730 $^0$C and we maintained a dynamic pressure of 0.03 mbar with the inlet flow rate of 3 sccm, consisting of 10\% ozone and 90\% molecular oxygen. We used a laser fluence about 1 Jcm$^{-2}$ with a frequency of 4 Hz. The target-substrate distance was 55 mm. After growth, the films were annealed at the flow rate of 10 sccm at 400 $^0$C for 15 minutes without pumping and then stopped heating until the films cooled below 55 $^0$C. And then pumped all the gas out and filled the chamber with 200 mbar molecular oxygen to anneal the films at around 200 $^0$C for 30 minutes. By the process, we can fill oxygen vacancy defects in La$_{2-x}$Sr$_x$CuO$_4$ films and remove excess oxygen induced by ozone. 

To enhance the scattering volume, we have synthesized superlattice structure for the interfacial superconducting films (Fig.~\ref{fig:characterization}(a)). The superlattice period comprises two unit-cell layers of La$_{2-x}$Sr$_x$CuO$_4$ (with M or S denoting overdoped metal $x$=0.45 or optimally doped superconductor $x$=0.15, respectively) and two unit-cell layers of insulating La$_2$CuO$_4$ (denoted as I). The nomenclature of the superlattice is dictated by the doping level, indicated as either 2M+2I or 2S+2I, and encompasses a total of 15 repeats within the superlattice structure. The metal La$_{2-x}$Sr$_x$CuO$_4$ ($x$=0.45) film is about 50 unit-cells and the superlattices are about 60 unit-cells. They are monitored by reflection high-energy electron diffraction (RHEED), indicating a layer-by-layer mode. And we can precisely control components and deposition order of layers in superlattices. The thickness of our 50 unit-cells La$_{1.55}$Sr$_{0.45}$CuO$_4$ film is 56 nm which is measured by X-ray Reflectivity (XRR). Other superlattice films are 82 nm and 73 nm for 2M+2I and 2S+2I. The XRR data were taken using a monochromatic Cu-K$\alpha$ source on a 3-kW high-resolution Rigaku Smartlab system. The film surfaces were examined using an atomic force microscope (AFM) \cite{SIinfo}, revealing their excellent condition with an average surface roughness (Rq) of 0.792 nm in a 50-unit-cell film. The Rq for the superlattice films was found to be 0.671 nm and 0.373 nm for 2M+2I and 2S+2I, respectively. The AFM data were collected using non-contact mode on a Park NX10 system. 

We investigated the temperature dependence of the resistance ($R$-$T$) of our superlattice films using a 4-point setup. To make electrical contacts, we deposited Ag film electrodes on the sample surface using electron-beam evaporation, which was then connected to Al wires using an ultrasonic wire bonding machine. The resistance was measured using a DC method in a commercial cryostat to obtain the R$_{sq}$(T) curve.

\subsection{Structural characterization and EELS measurements}

Scanning transmission electron microscopy (STEM) analysis was performed on a double-aberration corrected electron microscope (Thermo Fisher Scientific Titan Themis) with a probe forming semi-angle of 25 mrad and beam voltage of 300 kV. Energy dispersive X-ray spectroscopy (EDS) was used to map the distribution of each element. For EDS quantification, the background fitting and peak integration were performed after averaging 19 spectra to improve the sign-to-noise ratio. EELS measurements were performed using Gatan Quantum 965 spectrometer with an energy dispersion of 0.1 eV/channel. The energy resolution of EELS is $\sim$ 0.8 eV estimated from the full-width at half maximum of the zero-loss peak. Beam current of 30 pA was used for imaging and EELS to reduce beam damage effect. Before fitting the pre-edge peak of the O-$K$ edge, the background removed spectra were firstly averaged along the sample layer dimension to improve the signal-to-noise ratio. The intensity of the pre-edge O-$K$ is the area of the Gaussian peak fitted around the peak of 529 eV. 
Cross-sectional TEM sample was prepared using a Zeiss Auriga focused ion beam (FIB) and the surface damage layers were subsequently cleaned using a Fischione NanoMill.

\subsection{XAS and RIXS measurements}
The X-ray absorption (XAS) and RIXS were performed using the soft X-ray inelastic spectrometer at beamline 41A at the National Synchrotron Radiation Research Center (NSRRC) \cite{Taiwan41A}. The experiments were performed at base temperature $\sim$ $25$ K unless otherwise mentioned. We measured the XAS spectra at O $K$-edge with the total electron yield (TEY) method.
The resonant condition was achieved by tuning the energy of the incident X-ray to the Zhang-Rice singlet (ZRS) peak of the O $K$-edge ($\sim$ 529 eV). The RIXS spectra were collected with $\sigma$-incident polarization to maximize the charge signal. The reciprocal lattice units ($r.l.u.$) were determined in terms of the tetragonal unit cell with $a=b=3.78~\textrm{\AA}$ and $c=13.25~\textrm{\AA}$. We mainly used a grazing-out geometry, which by convention means positive $\mathbf{q_\parallel}$ (the projection of the transferred momentum along CuO$_2$ plane). The energy resolution was determined to be 25 meV. The acquisition time for each RIXS spectrum was 30 minutes (sum of 6 individual spectrum, each with 5 minutes) for the $\mathbf{q_\parallel}$ dependent measurements. All the spectra were normalized to the beam current I$_0$. 

\section{Results and discussion}

\subsection{Chemical composition and hole-doping distributions}

\begin{figure}[htbp]
\centering\includegraphics[width = \columnwidth]{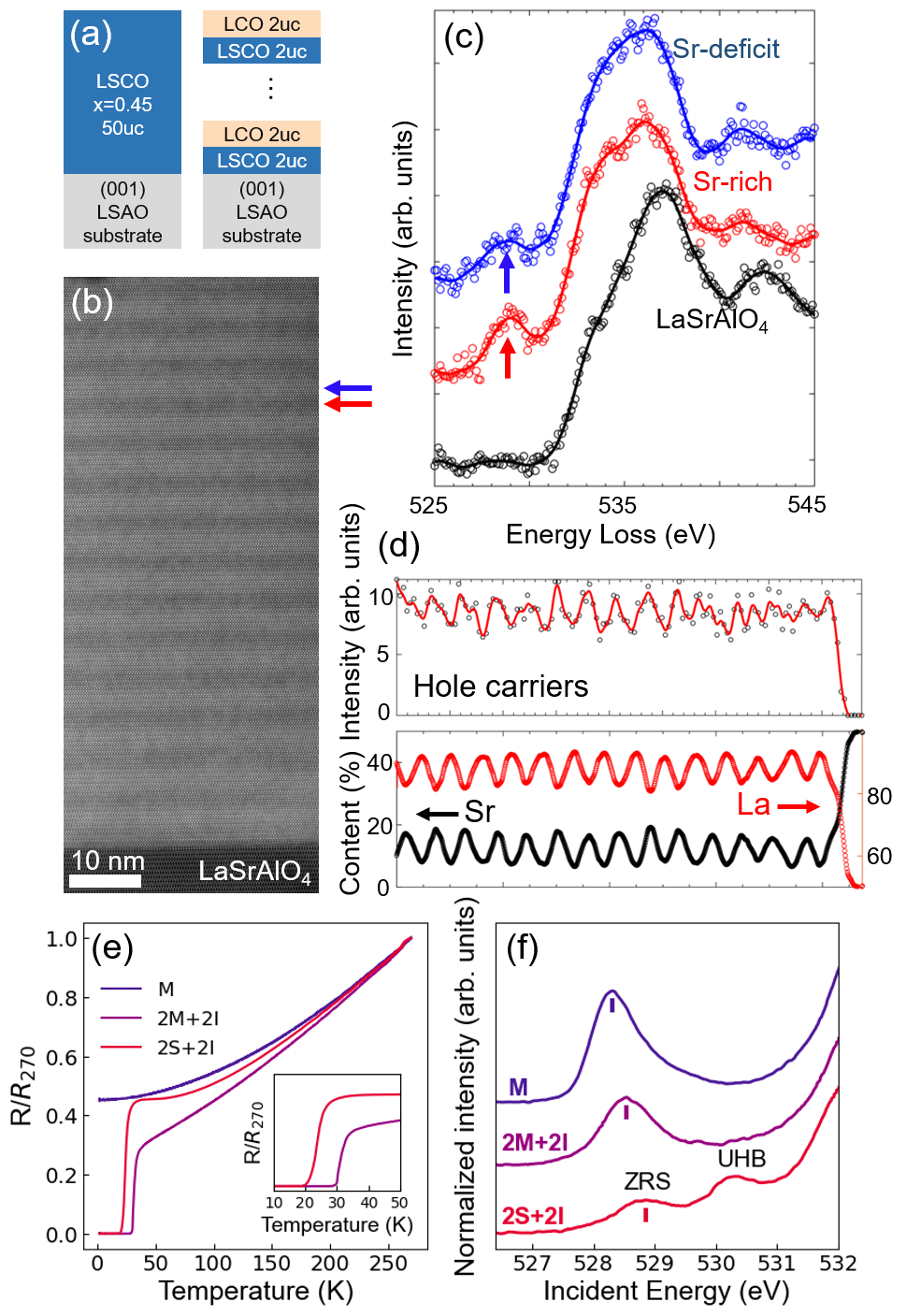}
\caption{\label{fig:characterization}\textbf{(a)} Schematic plots of single phase film (M) and superlattice (2S+2I, 2M+2I) deposited on LaSrAlO$_{4}$ substrate. Notation indicates: I is insulating La$_2$CuO$_4$; S is superconducting La$_{1.85}$Sr$_{0.15}$CuO$_4$; M is overdoped and metallic La$_{1.55}$Sr$_{0.45}$CuO$_4$. For superlattices, the first letter denotes the layer next to the LaSrAlO$_{4}$ substrate. \textbf{(b)} STEM high angle annular dark-field (HAADF) image of the 2M+2I superlattice. The Sr-deficit (La-rich) layers are brighter than the Sr-rich layers, which indicated by a blue and red arrow, respectively. \textbf{(c)} O $K$-edge EELS measurements on Sr-deficit and Sr-rich site, showing doping-dependent ZRS peak that is absent in the substrate. \textbf{(d)} The layer-dependent density of hole carriers (upper) and composition of La/Sr atoms (lower). \textbf{(e)} The dependence of resistance on temperature. The superconducting transition temperature is zoom-in in the inset. \textbf{(f)} The TEY XAS measurements of films, an offset is added for clarity. ZRS and Upper Hubbard Band (UHB) peaks have opposite tendencies with doping level \cite{NickXAS}.}
\end{figure}

The layer structure of the 2M+2I superlattice is shown in Figure~\ref{fig:characterization}(b) by using the cross-sectional view of scanning transmission electron microscopy (STEM) high angle annular dark-field (HAADF) image. The superlattice with alternative M and I layers is illustrated with bright (Sr-deficit) and dark (Sr-rich) contrast, since the STEM-HAADF image contrast is approximately proportional to $\sim$ Z$^{1.7}$ (Z is the atomic number) and the atomic number of La is higher than Sr. The sharp interfaces between M and I layers also suggest the high quality of the film. 
To further investigate the distribution of carriers along depth direction, we performed EELS measurements on the superlattice. Similar to XAS, the pre-peak at $\sim$ 529 eV is the ZRS peak originating from the hybridization between oxygen ligands and Cu $3d_{x^2-y^2}$ orbitals and its intensity can be used to estimate the hole density \cite{NickXAS,CTchenXAS}. The ZRS peak is noticeably stronger in the Sr-rich layers than in the Sr-deficit layers, while it is absent in the LaSrAlO$_4$ substrate (Fig.~\ref{fig:characterization}(c)). The hole density shows a clear periodic variation across the superlattice (Fig.~\ref{fig:characterization}(d)). 
We note that the average intensity ratio of the Sr-rich and Sr-deficit layers (M/I) is about 10/7, in stark contrast to the nominal Sr distributions, providing direct evidence for the charge redistribution in the superlattice. 
We also quantified the Sr and La concentrations using STEM-EDS \blue{(see Supplementary Fig. S2)}. There is an inverse variation in the content of La and Sr atoms along the normal direction, shown in Fig.~\ref{fig:characterization}(d). Specifically, in layer M, the maximum ratio of La and Sr is 4.3, i.e., Sr concentration of 0.38. Strong Sr diffusion occurs in layer I  resulting in a Sr concentration larger than 0.14. The absence of UHB in EELS in layer I (Fig.~\ref{fig:characterization}(c)) also verifies the doping level should be higher than 0.14 \cite{CTchenXAS}. The accurate doping level is difficult to obtain from O-K EELS due to the complicated experimental parameters, but it is clear that the doping level in layer M is higher than in layer I since the ZRS peak in O-K EELS is stronger. The tendency vs. the Sr concentration also agrees well with earlier XAS measurements in bulk LSCO single crystals \cite{NickXAS}. Therefore, the average doping of the 2M+2I superlattice can be predicted to be 0.24 based on the average Sr concentration estimated from EDS measurements.
It should be noted that the absolute value of the composition estimated from EDS can have a systematic error due to the strong multiple scattering in the zone-axis condition \cite{EDS_zhen}. Nevertheless, the trends of the Sr diffusion across the interfaces should be reliable. Therefore, the Sr profile across the M/I interfaces is much sharper than that of the hole density distribution, consistent with previous studies on LSCO/LCO superlattices \cite{dopingZn,LSCO/LCOREXS}.

Figure~\ref{fig:characterization}(e) shows the temperature dependence of the resistance (R-T), which reveals T$_c$ values of 30 K and 20 K for the 2M+2I and 2S+2I superlattices, respectively. The lower T$_c$ in the 2S+2I superlattice results from charge redistribution from the x=0.15 doped layer to the insulating layer. 
The R-T curves of the two superlattices resemble those of LSCO single crystals with dopings x=0.14 and x=0.075, respectively \blue{(see Supplementary Fig. S4)} \cite{SIinfo}.
The La$_{1.55}$Sr$_{0.45}$CuO$_{4}$ thin film exhibits metallic behavior.  We also performed O $K$-edge TEY XAS measurements on all three samples (Fig.~\ref{fig:characterization}(f)). The ZRS peak shifts to lower energy with an increase in hole doping, and our results for the 2M+2I and 2S+2I superlattices overlap well with those obtained for $x$=0.15 and $x$=0.07 single crystals \cite{CTchenXAS} \blue{(see Supplementary Fig. S4)} \cite{SIinfo}. Therefore, we defined the effective doping levels of our 2M+2I and 2S+2I superlattices to be $x_{eff}$=0.15$\pm$0.01 and $x_{eff}$=0.07$\pm$0.01, respectively.

\subsection{Charge order and phonon anomaly}

\begin{figure} [htbp]
\centering\includegraphics[width = \columnwidth]{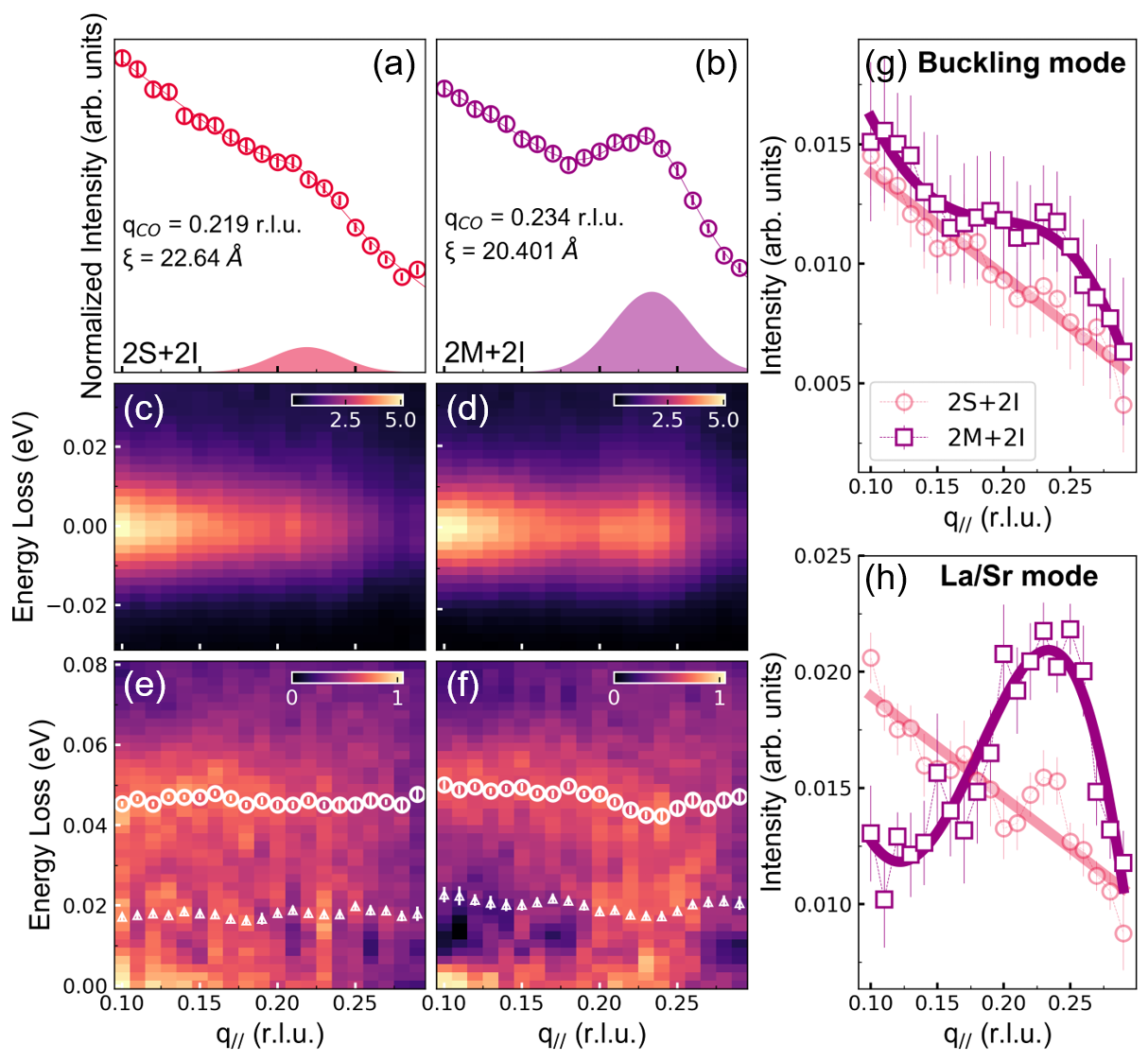}
\caption{\label{fig:CO in SL}
\textbf{(a,b)} Integrated intensity of the RIXS spectra measured at 25 K within the energy window [-0.02, 0.02] eV in 2S+2I and 2M+2I, respectively. The curves are fitted by a Gaussian function and a linear background. The wave vector (\textbf{q$_{CO}$}) and coherent length ($\xi$) are displayed in the figure. \textbf{(c,d)} RIXS map of the quasi-elastic region after normalization to orbital excitations. \textbf{(e,f)} RIXS map of the inelastic region after subtracting the elastic peak \blue{(see Supplementary Fig. S6)} \cite{SIinfo}. The energies of buckling and La/Sr phonon modes are marked by white circles and triangles. \textbf{(g,h)} Momentum dependence of the intensity for buckling mode and La/Sr phonon, respectively.}
\end{figure}

Charge orders are found ubiquitous in the cuprates family \cite{abbamonte2005spatially,Ghiringhelli2012,chang2012direct}, and phonons are observed softened at the \textbf{q$_{CO}$} by electron-phonon coupling and enhanced near the \textbf{q$_{CO}$} due to the interference between the dispersive charge excitations and the phonons \cite{dispersiveCDW,JieminCDW}. It is interesting to investigate how CO and phonons evolve in the interfacial superconducting superlattices. We performed RIXS measurements at the resonance of the ZRS peak and fixed the scattering angle at 150$^\circ$. 
 We have identified several excitations, including phonons (0.02 $\sim$ 0.07 eV), plasmon (0.1 $\sim$ 0.5 eV), bimagnons (0.6 $\sim$ 0.7 eV), and orbital excitations (1.5 $\sim$ 2.5 eV) \blue{(see Supplementary Fig. S5)} \cite{SIinfo}. 

\begin{figure} [htbp]
\centering\includegraphics[width = \columnwidth]{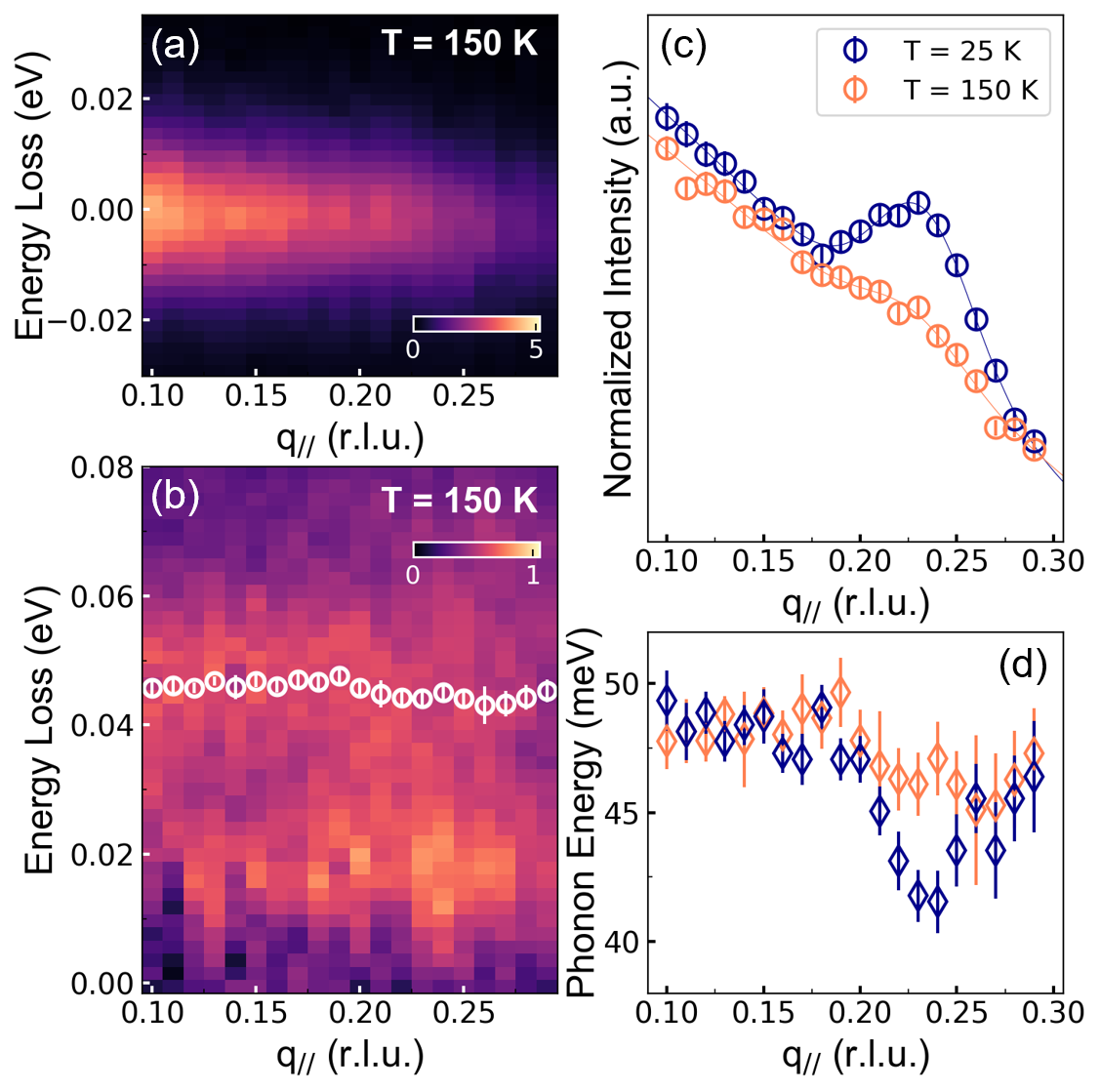}
\caption{\label{fig:COTdep} 
\textbf{(a,b)} RIXS map of the quasi-elastic region and phonons after subtracting the elastic peak in 2M+2I superlattice measured at 150 K, respectively. Buckling phonon modes are marked by white circles. \textbf{(c)} Integrated intensity of the RIXS spectra within the energy window [-0.02, 0.02] eV at 25 K and 150 K. The curves are fitted by a Gaussian function and a linear background. \blue{(see Supplementary Fig. S8)} \cite{SIinfo}. \textbf{(d)} The dispersion of buckling phonon modes measured at 25 K and 150 K.}
\end{figure}

Here we focus on the CO and phonons in the low-energy region. As illustrated in Fig.~\ref{fig:CO in SL}(a-d), the CO was detected in both 2S+2I and 2M+2I superlattices. In 2M+2I, the CO wave vector \textbf{q$_{CO}$} = $(0.23,0)$ and coherent length $\xi$ $\sim$ 5.4a are similar to those observed in optimal doped LSCO single crystal ($x$=0.16) \cite{LSCOSDWCDW}. While the CO is weaker in 2S+2I due to its lower doping level, we have nonetheless identified it at \textbf{q$_{CO}$} = $(0.22,0)$. The decrease of \textbf{q$_{CO}$} with lower effective doping in the superlattice is consistent with the doping dependence of \textbf{q$_{CO}$} in La-based cuprates \cite{LSCOqCDW}. To examine the phonon response, we subtracted the elastic peaks and generated the maps in Fig.~\ref{fig:CO in SL}(e,f). We have observed three phonon branches at 18 meV, 45 meV and 75 meV, which can be assigned to the La/Sr vibration mode \cite{Taiwan41A}, bond-buckling and bond-breathing phonon modes, respectively. These findings are in accordance with a previous RIXS study on optimally doped LSCO \cite{huangCDF}. In the 2M+2I superlattice, we find that the energies of the La/Sr vibration and buckling phonon branches exhibit softening at \textbf{q$_{CO}$} (Fig.~\ref{fig:CO in SL}(e,f)), and the phonon intensities increase near \textbf{q$_{CO}$} at 25 K (Fig.~\ref{fig:CO in SL}(g,h)). Moreover, both CO and phonon anomaly in 2M+2I weaken at 150 K (Fig.~\ref{fig:COTdep}(c,d)). This agrees with the temperature-dependent behavior of CO below the doping level of $x$=0.16 in LSCO \cite{FateCO}, beyond which the CO changes into temperature-independent fluctuations. Thus, the behavior of CO in 2M+2I aligns well with an effective doping of $x_{eff}$=0.15$\pm$0.01. In contrast, in the 2S+2I superlattice, the weak CO signal does not result in observable phonon softening and intensity anomaly (Fig.~\ref{fig:CO in SL}(e,g,h)).

Our measurements reveal several noteworthy findings. Firstly, the CO and phonon softening persist in the reduced dimensionality of the interface. Secondly, our observation of the 2M+2I superlattice is in stark contrast to its individual constituents, La$_{2}$CuO$_{4}$ (I) and La$_{1.55}$Sr$_{0.45}$CuO$_{4}$ (M). Specifically, the former does not exhibit a CO, while the latter shows a re-entrant CO with \textbf{q$_{CO}$} at $(0.166,0)$ that remains nearly temperature-independent up to room temperature \cite{ODLSCOCO}. 
It was anticipated that the width of the CO peak would broaden due to the shifting of \textbf{q$_{CO}$} with varying dopings within the superlattice. However, we have observed that the width of the charge order (CO) peak in the 2M+2I superlattice closely resembles that of CO with an effective doping level of $x_{eff}$=0.15$\pm$0.01, with no apparent broadening. Finally, our results demonstrate that the CO in LSCO emerges as early as $x_{eff}$=0.07$\pm$0.01, extending its presence to a more underdoped region. It has been proposed that the $T_c$ enhancement in the interfacial superconductor LSCO/LCO, when compared to bulk LSCO, is due to the suppression of some competing instability via the long-range strain or electrostatic effects \cite{interfacialSC}. However, our observation of charge order in LSCO/LCO does not support this argument for the enhancement of $T_c$.

\subsection{Plasmon}

In bulk cuprates plasmon is a three-dimensional dispersive charge excitation originating from the interaction between neighboring CuO$_2$ planes.  
The plasmon exhibits a linear dispersion at small in-plane momentum transfer (\textbf{q$_{||}$}) and a periodic dispersion along the out-of-plane direction, which has been observed by recent RIXS experiments \cite{LCCOplasma,plasmonZhouKJ}. Here we investigate the evolution of plasmon from bulk to interface samples and determine whether the periodically distributed charge along the depth direction affects the interaction between the CuO$_2$ planes.

\begin{figure} [htbp]
\centering\includegraphics[width = 0.92\columnwidth]{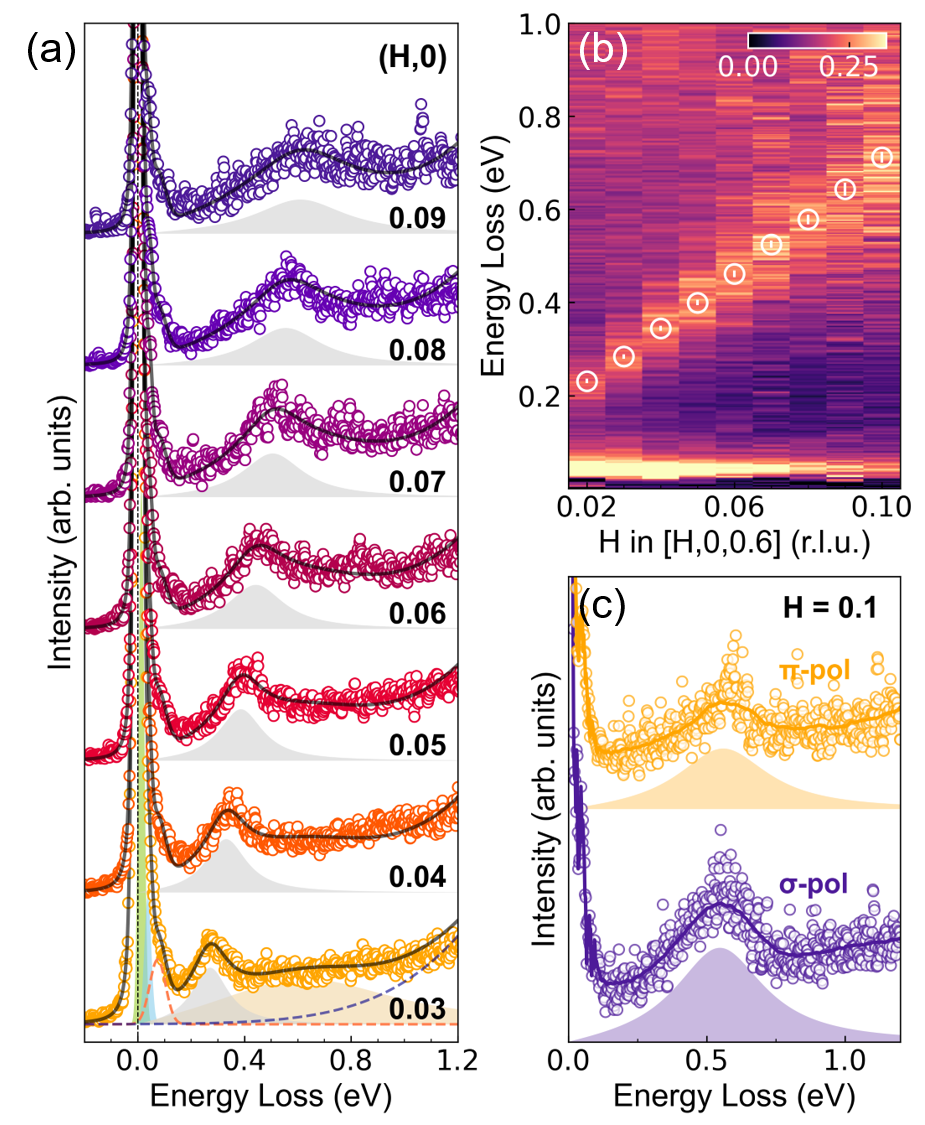}
\caption{\label{fig:plasmon in LSCO(0.45)}\textbf{(a)} Observation of plasmon in overdoped metallic La$_{1.55}$Sr$_{0.45}$CuO$_{4}$ film with fixed $L$ $\sim$ 0.6 $r.l.u.$. Fits with different components, including phonons (green-, blue-shaded and orange dashed lines), plasmon (grey-shaded), bimagnons (orange-shaded), and orbital excitations (black dashed line). An offset is added for clarity. \textbf{(b)} RIXS map of plasmon in La$_{1.55}$Sr$_{0.45}$CuO$_{4}$ film along $(H,0)$ direction. The map was plotted after normalization to orbital excitations. \textbf{(c)} Polarization dependence of plasmon, which is stronger at $\sigma$-polarization. }
\end{figure}

We first study the plasmon in the heavily overdoped metallic La$_{1.55}$Sr$_{0.45}$CuO$_{4}$ film. We fixed the $L$ at 0.6 $r.l.u.$ by changing both the incident angle and scattering angle when \textbf{q$_{||}$} changed \cite{LCCOplasma}. As shown in Fig.~\ref{fig:plasmon in LSCO(0.45)}(a,b), we observe a linear dispersive excitation below 0.8 eV and the intensity is weaker at $\pi$-pol than at $\sigma$-pol (Fig.~\ref{fig:plasmon in LSCO(0.45)}(c)). This corresponds to the charge scattering signal and is consistent with the previous results of plasmon \cite{LCCOplasmondoping,nphysplasmon}. We extracted the plasmon (shaded grey area in Fig.~\ref{fig:plasmon in LSCO(0.45)}(a)) following a standard fitting process \blue{(see Supplementary Fig. S7)} \cite{SIinfo}. The group velocity of the plasmon is $\sim$ $3.8 \pm 0.2$ $eV \AA$, similar to the RIXS results in underdoped bulk LSCO \cite{plasmonZhouKJ}.  
Our observation of plasmon in overdoped metallic LSCO extends the plasmon doping range beyond the SC phase.

\begin{figure}[htbp]
\centering\includegraphics[width = 0.92\columnwidth]{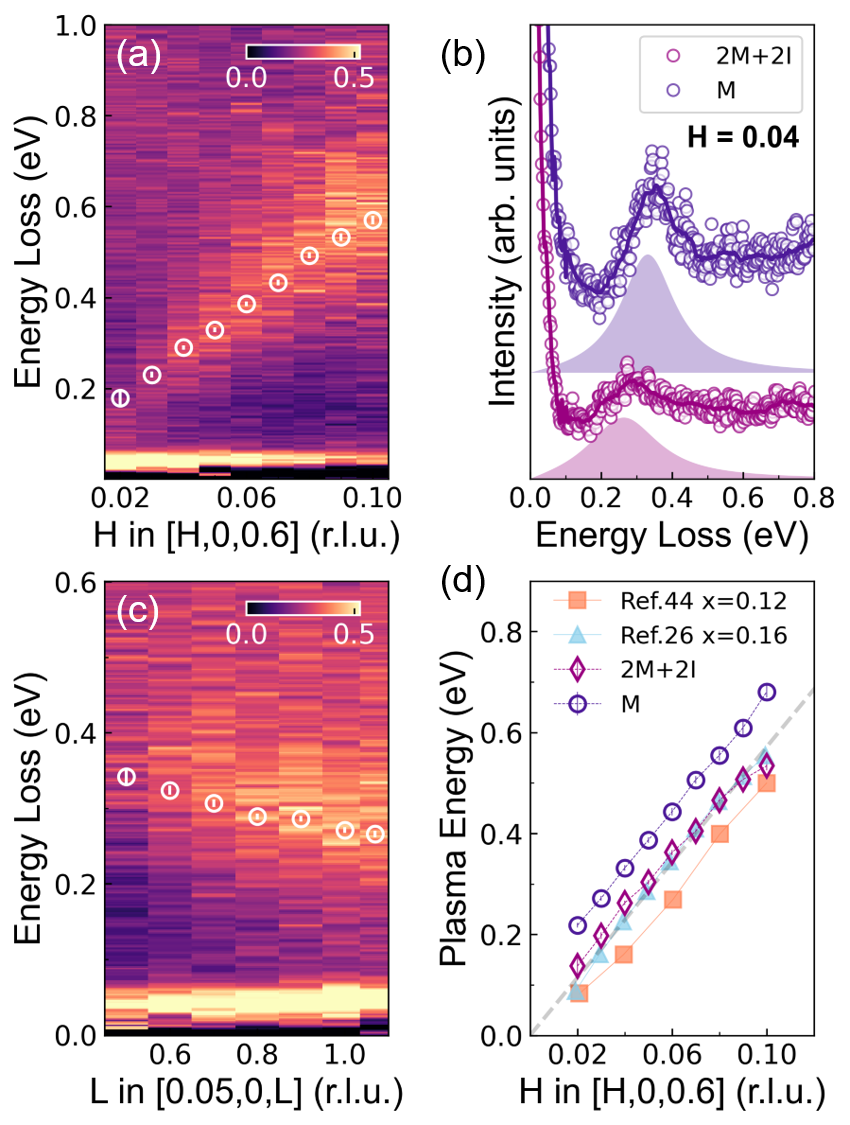}
\caption{\label{fig:plasmon in SL}\textbf{(a)} Observation of plasmon in 2M+2I superlattice at fixed $L$ $\sim$ 0.6 $r.l.u.$ The maps were plotted after normalization to orbital excitations. \textbf{(b)} Comparison between plasmon in M film and 2M+2I superlattice. An offset is added for clarity. \textbf{(c)} The $L$ dependence of plasmon at fixed $H$ $\sim$ 0.05 $r.l.u.$ (\blue{see Supplementary Fig. S7} for fitting details \cite{SIinfo}). \textbf{(d)} The in-plane dispersion of plasmon. Data of single crystals are cited from reference \cite{Huangplasmon,plasmonZhouKJ}.}
\end{figure}

We then pay attention to the plasmon in the 2M+2I superlattice. As shown in Fig.~\ref{fig:plasmon in SL}(a), we observe a similar linear dispersion along the $(H,0)$ direction. We compare the plasmon spectra of M and 2M+2I in Fig.~\ref{fig:plasmon in SL}(b). The plasmon peak is more pronounced and has higher energy in M than in 2M+2I. We summarize the extracted plasmon energy and compare them with those of LSCO single crystals (Fig.~\ref{fig:plasmon in SL}(d)) \cite{Huangplasmon,plasmonZhouKJ}. The energy of the plasmon in 2M+2I agrees well with that of $x$=0.16 single crystal \cite{plasmonZhouKJ}.  
Moreover, we observed that the plasmon in LSCO shifts to higher energy with increasing hole doping levels up to x=0.45. In electron-doped cuprate La$_{2-x}$Ce$_x$CuO$_4$ (LCCO), the plasmon energies are proportional to the square of doping ($\sqrt{x}$) up to $x$=0.15 and saturate at higher dopings for $x$=0.18, which may relate to Fermi-surface reconstruction \cite{LCCOplasma}. The plasmon gap is approximately zero for the 2M+2I superlattice. On the other hand, the plasmon in the M film has finite energy of $\sim$ 0.1 eV when extrapolated to the \textbf{q$_{||}$}=0 point, which is reminiscent of the plasmon gap in the electron-doped cuprate superconductor Sr$_{0.9}$La$_{0.1}$CuO$_2$ (SLCO) ($\sim$ 0.12 eV) \cite{Gappedplasmon}. Their study proposed that the gap was related to the nonzero interlayer hopping (t$_z$) and could be analyzed quantitatively using the t-J-V model. 
We estimate that the interlayer hopping in the M film is t$_z$ $\sim$ 0.05 eV, about half the gap at the \textbf{q$_{||}$}=0 point following their method. 
The interlayer hopping in infinite layer SLCO cuprates is enhanced due to the closely spaced adjacent CuO$_2$ planes, while the high doping level in overdoped LSCO may also increase the transition probability between CuO$_2$ layers. 

To study the effects of periodic electron distribution along the out-of-plane direction, we also measured the $L$ dependence of the plasmon at fixed $H$ = 0.05 $r.l.u.$  in the 2M+2I superlattice (Fig.~\ref{fig:plasmon in SL}(c)). Due to the limited reciprocal space accessible at the O $K$-edge, we could only reach $L$ up to $\sim$ 1 $r.l.u.$ and observed a local minimum of the plasmon energy there, which is similar to the plasmon in bulk LSCO \cite{LCCOplasma,plasmonZhouKJ}. This suggests that the periodic electron distribution of the superlattice does not affect the out-of-plane plasmon dispersion. Finally, we increased the temperature to 150 K and found that the plasmon was nearly temperature-independent \blue{(see Supplementary Fig. S9)} \cite{SIinfo}, consistent with a recent study on LSCO thin films \cite{Hepting_2023}. 
In contrast, the plasmon in electron-doped cuprate Nd$_{2-x}$Ce$_x$CuO$_4$ showed weak temperature dependence at $x$=0.147 and strong temperature dependence at $x$=0.166, which was proposed to associate with a symmetry-broken state  \cite{nphysplasmon}. The different temperature behavior of plasmon in hole-doped and electron-doped cuprates remains to be understood.

\section{Conclusion}

In summary, our atomic resolution EELS and EDS measurements offer direct evidence of charge redistribution that deviates from the Sr diffusion profile across different layers of the superlattice. Additionally, our RIXS measurements reveal that the behaviors of charge order, phonons, and plasmon in the superlattices align with the effective dopings determined from interfacial superconductivity, rather than corresponding to the properties of its individual constituents or the average doping.
These findings underscore the pivotal role of effective doping in influencing both the transport properties and elementary excitations within LSCO/LCO superlattices. However, the mechanism behind the enhancement of T$_c$ in interfacial LSCO/LCO superconductors remains an outstanding question. Future investigations with improved energy resolution in EELS and RIXS may illuminate this issue further by delving into the study of phonons at the interface.\\

\vspace{1 ex}
\begin{acknowledgments}
\noindent
The RIXS experimental data were collected at beamline 41A of the National Synchrotron Radiation Research Center (NSRRC) in Hsinchu 30076, Taiwan. Y. Y. P. is grateful for financial support from the Ministry of Science and Technology of China (Grants No. 2019YFA0308401 and No. 2021YFA1401903) and the National Natural Science Foundation of China (Grant No. 11974029). Z. Chen acknowledges the financial support from the National Natural Science Foundation of China (Grant No. 52273227). Y. Xie acknowledges the financial support from the National Natural Science Foundation of China (Grant No. 12074334).
\end{acknowledgments}

\end{document}